# Smart Contracts for SMEs and Large Companies

Christian Gang Liu
Faculty of Computer Science
*Dalhousie University*
Halifax, Canada
Chris.Liu@dal.ca

Peter Bodorik
Faculty of Computer Science
*Dalhousie University*
Halifax, Canada
Peter.Bodorik@dal.ca

Dawn Jutla
Sobey School of Business
*Saint Mary's University*
Halifax, Canada
Dawn.Jutla@gmail.com

*Abstract*— Research on blockchains addresses multiple issues, with one being writing smart contracts. In our previous research we described methodology and a tool to generate, in automated fashion, smart contracts from BPMN models. The generated smart contracts provide support for multi-step transactions that facilitate repair/upgrade of smart contracts. In this paper we show how the approach is used to support collaborations via smart contracts for companies ranging from SMEs with little IT capabilities to companies with IT using blockchain smart contracts. Furthermore, we also show how the approach is used for certain applications to generate smart contracts by a BPMN modeler who does not need any knowledge of blockchain technology or smart contract development - thus we are hoping to facilitate democratization of smart contracts and blockchain technology.

## I. Introduction

The publication of the Bitcoin white paper in 2008 and the subsequent launch of the Bitcoin blockchain in 2009 have catalyzed extensive interest and research into blockchain technology. This technology has attracted widespread attention from businesses, researchers, and the software industry due to its compelling attributes, such as trust, immutability, availability, and transparency. However, as with any emerging technology, blockchains and their associated smart contracts present new challenges, particularly in areas such as blockchain infrastructure and smart contract development.

Ongoing research is actively addressing several critical issues, including blockchain scalability, transaction throughput, and the high costs associated with consensus algorithms. Additionally, smart contract development faces unique difficulties, such as limited stack space, the oracle problem, data privacy concerns, and cross-blockchain interoperability. These topics have been explored in-depth, with numerous comprehensive literature reviews available [e.g., 1, 2].

The constraints imposed by blockchain technology increase the complexity of smart contract development, which is well documented in various literature surveys, such as [3, 4]. To address these difficult challenges and simplify smart contract development, researchers such as López-Pintado et al. (2019) [5, 6], Tran et al. (2018) [7], Mendling et al. (2018) [8], and Loukil et al. (2021) [9] have proposed using Business Process Model and Notation (BPMN) models that can be transformed into smart contracts.

We also use BPMN modeling to represent the application requirements, but we use a different approach to transform BPMN models to smart contracts. Instead of transforming the BPMN models directly to smart contract methods, we exploit multi-modal modeling to represent the flow of computation of the business logic in a blockchain-independent manner. To show the proof of concept, we developed a tool, **T**ransforming **A**utomatically **B**PMN model into **S**mart contracts, called **TABS**, to generate smart contract from BPMN models while also supporting side-chain processing [10].

In [11] we extended the TABS tool and its underlying concepts into a tool TABS+ that allows representing multi-step activities of actors using nested trade transactions while also providing, in automated fashion, supporting mechanisms to enforce the transactional properties [11] of the nested multi-step transactions.

Most recently, we further extended the underlying concepts and the tool to support upgrade/repair of smart contract, which is necessary (i) to repair bugs in smart contracts and/or (ii) to amend the smart contracts to model new functionalities or features in business processes as they continually evolve [12].

One of our main objectives is to automate generation of smart contracts from BPMN models such that the transformation process can be managed by a BPMN modeler without (much) intervention by IT support with expertise on blockchains smart contracts. Although our approach has brought us closer to that objective, services of a software developer are still required to write some well-defined methods for the BPMN task elements.

### A. Objectives and Contributions

We have two objectives achieving of which also form the paper's contributions. Our first objective is to show that, for certain types of blockchain applications, our approach can generate smart contracts in automated fashion from BPMN models without assistance of a software developer. Although this limits the type of applications that can be supported, the benefit that is gained is the generation and deployment of smart contracts directly from BPMN models that can be exploited by organizations without the usual support of developers of smart contracts.

Our second objective is to show that our approach can be used to support generation of smart contracts from BPMN models under various scenarios ranging from use by SMEs to use by large companies with sophisticated IT infrastructure that also utilizes blockchains to support its internal activities as well as collaborations with partner organizations.

*B. Outline*

The second section provides background. The third section describes how we are augmenting our approach and the tool to support generation of smart contracts without the need of a software developer, albeit for a subset of BPMN models that satisfy certain conditions. The fourth section describes how our approach is suitable for use by SMEs as well as by large companies. The fifth section provides related work, while the last section provides summary and conclusions.

## II. BACKGROUND

We overview BPMN modeling first, then the use of Hierarchical State Machines (HSMs) and multi-modal modeling in system analyses, and then our approach to generating smart contracts from BPMN models.

*A. Business Process Management Notation (BPMN)*

Business Process Model and Notation (BPMN), developed by the Object Management Group (OMG) [13-16], is a standard that was designed to be accessible to a diverse range of business users, including analysts, technical developers, and managers. The widespread practical adoption of BPMN is evidenced by the variety of software platforms that facilitate the modeling of business processes with the aim of automatically generating executable applications from BPMN models. For instance, the Camunda platform converts BPMN models into Java applications [17], while Oracle Corporation translates BPMN models into executable process blueprints using the Business Process Execution Language (BPEL) [18].

BPMN models are characterized by several key features, including flow elements that represent the computational flows between different BPMN components. A task within a BPMN model signifies computation that is executed when the flow reaches the task element. Other elements in BPMN manage the conditional branching and merging of computational flows, with Boolean expressions (guards) used to control the flow of computation. Furthermore, BPMN also models various events that may arise and how these events are caught and processed. Additionally, data elements within BPMN models describe the data or objects that move along with the computations, serving as inputs for decision-making in guards or computation tasks.

*B. FSMs, Hierarchical State Machines (HSMs), and Multi-modal Modeling*

Finite State Machine (FSM) modeling has been extensively utilized in software design and implementation, often enhanced with features such as guards on FSM transitions. In the late 1980s, FSMs evolved into Hierarchical State Machines (HSMs), in which a state in an FSM can be represented by an FSM itself. Although HSMs do not increase expressiveness of FSMs, they lead to hierarchical FSM structures to facilitate the reuse of patterns by allowing states to contain nested FSMs [19].

Girault et al. (1999) [20] explored the combination of HSM modeling with concurrency semantics derived from models like Communicating Sequential Processes [21] and Discrete Event systems [22]. They demonstrated that a system state could be represented by an HSM, where a specific concurrency model is applied exclusively to that state. This approach enables multi-modal modeling, allowing different hierarchical states to employ the most appropriate concurrency models for the concurrent activities within those states. We exploit multi-modal modeling to express the flow of computation within a BPMN model in a blockchain-agnostic way by using DE modeling to represent concurrency while concurrent FSMs are used to express functionality.

*C. BPMN Model Transformation to Smart Contract Methods and TABS+R Tool*

In [10], we presented a methodology for transforming BPMN models into smart contracts. The transformation process involves several key steps:

1. *Transformation to DE-HSM Model:* The BPMN model is first transformed into a graph representation and then into a DE-HSM model.

2. *Analysis and multi-step trade-transaction specification:* The model's computation flow is analyzed to identify localized sub-graphs that are then used to define nested, multi-step trade transactions.

3. *Transformation to DE-FSM Model:* The DE-HSM model is elaborated by recursive replacement of each DE-HSM model with its elaborated DE-FSM model and thus flattening the entire DE-HSM model into an interconnected network of DE-FSM (Discrete Event Finite State Machine) sub-models.

4. *Transformation to Smart Contracts:* The interconnected DE-FSM models are transformed into smart contract code.

It should be noted that the flow of computation in the smart contracts is represented by DE modeling combined with functionality represented by concurrent FSMs – and these are blockchain independent. As long as the target blockchain has a smart contract deployed containing the TABS monitor, any smart contract generated by the transformation process can be deployed and executed on that target blockchain. The monitor smart contract provides the execution environment for the DE modeling and concurrent FSMs. In short, the monitor has a detailed view of the business logic flow, including the corresponding data flowing along with the flow of computation, wherein the business logic is expressed in an abstract manner, using DE modeling techniques and concurrent FSMs, and is thus blockchain independent.

## III. ATTESTATION FOR AUTOMATED GENERATION

One of our objectives is to achieve generation of smart contracts that are blockchain agnostic. We made progress towards this objective by representing the flow of collaboration logic in a blockchain-independent as described above. However, currently, the scripts for the BPMN task elements need to be coded/written by a software developer in a specific computer language executable by the target blockchain.

To overcome this issue of the dependence on coding of task elements, in this section we describe how we adapted a two-layer approach taken by the Plasma project, described in [23], to generate smart contracts without writing scripts for the BPMN task elements. The Plasma project approach to improve scalability uses two chain layers, wherein the sub-servient chain performs the transaction detailed work, while the main chain records the certifications of the results of work performed by the subservient chain, such as a sidechain. This approach was used for scalability by the Ethereum public blockchain [24] in that the main Beacon Chain simply records coordination activities in managing the consensus and approvals of blocks appended to shards and in storing results of attestation of shard blocks.

We utilize a similar approach in that the scripts of the BPMN task elements are executed off the mainchain, while the smart contract executed on the mainchain simply guides the collaborations and obtains certifications about the results of the tasks executed off chain.

*A. Motivation for Certifications of Work of Task Elements*

The BPMN task element represents computation, within a swimlane (BPMN terminology) of one actor, on data flowing into the task element. The task uses the data flowing into the computation and the content of state variables to produces data flowing out of the computation while also updating state variables. For some applications, the task element examines the details of a document flowing in and makes decisions based on the data contained within that document. By having such computation performed by a smart contract, trust is achieved as all parties can examine details.

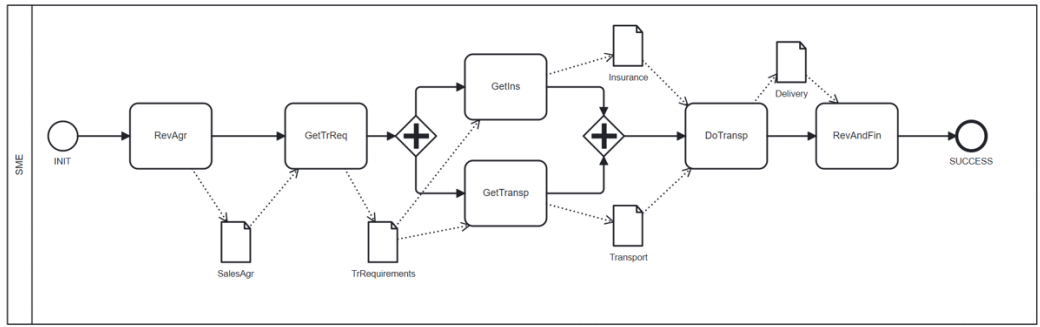

Fig. 1. PMN Model for a Sale of Product and Its Delivery

However, such computation also causes difficulties due to amendments required for either repairing bugs or for new features that need to be introduced, as it is likely that the required amendments would be within the task elements that are executed as a part of a smart contract. And repairing/upgrading smart contracts is not easy [25-27].

Many applications include simpler interactions amongst partners/actors, interactions that consist of exchange of documents rather than performing computations on such documents. In such situations, task elements need not be used, and instead we use prepared interactions for certified exchange of documents.

We examined sample use cases appearing in the literature, use cases detailing transformation of BPMN models into smart contracts, with examples being Order-Supply [28], Supply Chain [29], Parts Order [30], Sales and Shipment [31], and Ordering Medications [32]. In all of them, besides transferring documents amongst actors, the document creation, review, or amendment are performed off-chain by a single actor. In fact, for some use cases, such as the case of the supply chain management [32], data exchanged between the actors only consists of exchanging QR codes identifying documents that are exchanged – the smart contract interaction between the partners is in terms of documents being exchanged.

Thus, if the task method execution can be performed off-chain, then the code for the task script element does not need to be provided as long as the generation of the smart contracts from BPMN model facilitates certified exchange of documents between the on-chain and off-chain computation.

For exposition purposes, we are going to use a simple BPMN model, shown in Fig. 1, for a sale of a large product, such as a combine harvester. The model shows that an agreement on the sale of the product is reached first, which is followed by arrangements for the transport of the product. Transport arrangements include finding the requirements for the transport of the product, such as or safety requirements in case of dangerous products in transport. Once the transport requirements are determined, the insurance and transport are arranged, and the product is shipped/transported. Following the transport, the product is received, and payments are completed.

*B. Certification of Exchanged Documents*

Recall that as part of BPMN modeling, the modeler is asked to use data association elements to describe the purpose of the task and describe the data/information flowing along with the flow of computation, and hence also flowing in and out of the task element. This information is also passed to the off-chain component together with a document that is input (flows) into the task element. Once the task is completed, output from the task element is a document that is passed along the flow of computation.

As is the usual practice for blockchains, a document is stored off-chain, while it is the digitally signed hash-code of the document that is stored on the blockchain, wherein the signed hash code is used to confirm the document authenticity, where the authenticity includes confirmation of (i) authorship and (ii) that the document has not been modified.

For storage of documents, we currently utilize the Inter Planetary File System (IPFS) [33]. When a document is created, uploading it to IPFS generates a new content-addressed hash code identifier (CID), which is signed and stored by the smart contract. This allows the on-chain components associated with BPMN data elements to interact with the off-chain document without needing to directly handle its content.

For example, the first task receives a purchase offer document from an external source. An accepted purchase offer results in a sales agreement that is used in subsequent processing. The sales agreement is represented by an association data element, *SalesAgr*. The dotted arrow from *RecAgr* to the association data element SalesAgr signifies the creation of the *SalesAgr* by the *RecAgr* task. The dotted arrow from the *SalesAgr* by the *GetTrReq* task element signifies that the *SalesAgr* is delivered for further processing to the *GetTrReq* task.

The *GetTrReq* task determines the transport requirements for the product and stores them in a newly created IPFS document *TrRequirements*. The CID of the document is forwarded to next step in processing. The transport requirements are forwarded to the *GetIns* and *GetTransp* task elements that can be executed concurrently as shown by the fork gate represented by a diamond with a plus sign in it. The *getIns* task produces the insurance contract, called *Insurance*, while the *getTransp* creates a *Transport* document that is a contract for the transport of the product. Once the insurance and the transport contracts are obtained and provided to the transporter, the product be delivered to the destination, which is represented by the task *DoTrasp*. Completed delivery is documented in the document called *Delivery* that is forwarded to the final task, *RecAndFin*, to indicate reception of the product by the purchaser and finalization of the contract.

It should be noted that for brevity only a simplified model was presented that ignores many details, such as not accepting the purchase offer, deposits or final payments.

This standard interaction model for storing documents off-chain is used to prevent the blockchain from being overburdened, while still allowing transactions to be secure and complex multi-step processes to be executed. Additionally, any update or modification to a document generates its new CIDs, effectively handling version control and verification throughout the smart contract's lifecycle.

Thus, for applications that include collaborations that involve exchange of documents, the computations associated with the task elements can be off-loaded to off-chain resulting in facilitation of generation of smart contracts without requiring scripts for the BPMN task elements. Under such circumstances, our approach and tool to generation of smart contracts from BPMN models can be automated without intervention of a software developer and can be under the control of a Business Analyst (BA) who develops the BPMN model and asks the tool (i) to transform it into smart contract for the target blockchain and that (ii) to deploy the smart contract on the target blockchain.

In short, when the work of a task element can be executed off-chain and the interaction between the on-chain and off-chain components can be modeled simply by a certified exchange of documents, then the transformation of the BPMN model into a smart contract is used to support such a certified exchange of documents and thus avoid coding of the task elements. Consequently, a BPMN model can be transformed into a smart contract in automated fashion and deployed on the target blockchain under the control of the BA without assistance of a software developer.

Currently, we support certified information exchange between the on-chain and off-chain components using HTTP web services. As an example, consider the communication between the seller company and the insurance company. First, the on-chain component generates a request to the insurance company to get the insurance while providing it with a document containing the product description and transport requirements. The insurance company responds either with a negative response or with a positive response providing the seller with the insurance contract. As smart contracts are not able to access external resources, the smart contract raises an event that is captured and results in servicing the event by invoking the web service requesting insurance. The web service will produce the insurance document and will provide it to the mainchain smart contract by a call to a smart contract method to receive the insurance document/contract, wherein such a certified exchange is generated by the transformation process of the BPMN model into smart contract.

## IV. SMART CONTRACTS FOR SMEs AND LARGE COMPANIES

To show the flexibility of our approach, we are going to utilize the example use case, shown in Fig. 1, under two different scenarios, one in the context of a small SME, while the other one in the context of a large organization with sophisticated IT department that has expertise on writing smart contracts.

### A. Use Case in the Context of an SME

An SME would like to use a smart contract to ensure secure computation and obtain certified documentation on the trade activity. An SME's Business Analyst (BA), who is familiar with BPMN modeling, uses the TABS+R tool to create a BPMN model shown in Fig. 1. The BA creates the BPMN model and specifies that the task elements are executed off-chain and that the system should facilitate exchange of documents between the smart contract and the off-chain computation.

For an SME, off-chain computation may simply be manual by, perhaps, BA performing the off-chain work. For instance, for the *GetTrReq* task, the BA may contact a registry and find the transport requirements and store them in a newly created IPFS document *TrRequirements*. The CID of the document is forwarded to next step in processing. The transport requirements are forwarded to the *GetIns* and *GetTransp* task elements that can be executed concurrently as shown by the fork gate represented by a diamond with a plus in it. The BA may communicate with the insurance company for an insurance contract represented by the *Insurance* document that is stored on IPFS. Similarly, BA may negotiate a contract for the transport of the product, wherein the transport contract is stored in the *Transport* document on the IPFS. Once the insurance and the transport contract are obtained, they are forwarded to the *DoTransport* task. Once the product is delivered, the transporter returns a document, called *Delivery* that contains information on the delivery of the product. The *Delivery* document is forwarded to the *RecAndFin* to receive the *Delivery* document and finalize the trade activity.

For the *DoTransport* task, the insurance and the transport agreement would be input into the off-chain task, wherein the transporter would perform the transport and at the completion of the task would provide a document with confirmation of the product's arrival at the destination. The smart contract records the activities performed by the BA while storing the documents on IPFS with their CIDs stored on the blockchain smart contract.

There is some initial setup required before an SME can create smart contracts from BPMN models. The SME's target blockchain would need to be identified so that the generated smart contract can be deployed on the target blockchain. Furthermore, initially, the smart contract containing the TABS+R monitor would need to be deployed on the blockchain. However, this is only a one-time initial overhead that is amortized over all smart contracts generated by the approach for that target blockchain. Furthermore, this task is also automated as it simply involves deploying the TABS+R monitor smart contract on the target blockchain. Currently, we provide the TABS+R monitor smart contracts for Hyperledger Fabric (HLF) and for blockchains based on Ethereum Virtual Machine (EVM).

*B. Use Case in the Context of a Large Company*

Assume now that a similar application is being developed in the context of a large company with sophisticated IT systems. The company now has two departments, one for sales and one for the product shipment, and uses cutting-edge technologies, such as blockchains for collaborations and AI for automation. Once the sales agreement has been reached by the sales department, the sales agreement, which includes the product description and the purchaser information, needs to be communicated to a shipment department that uses its own internal processes to facilitate the product shipment to the purchaser.

A BPMN model that may represent the application is shown in Fig. 2. However, as showing the creation and exchange of documents would clutter the figure, we do not show exchange of such documents explicitly.

In comparison to Fig. 1, Fig. 2 has significant differences as information is flowing across departments and external actors that include the buyer, insurance company, transport company, and the registry of transport requirements. In BPMN, actor activities are contained in a swimlane that is represented by a rectangle. Information flow between actors is represented by lines that cross swimlanes. Thus, instead of a single swimlane as shown in Fig. 1, there are multiple swimlanes in Fig 2. There is a swimlane for each of the company's sales and shipping departments, denoted as *SalesDep* and *ShipDep*, respectively; and a swimlane for each of the external actors that include the buyer, the transport-requirements registry (*ReqRegistry*), insurance company (*InsComp*), and the transporter (*Trnasp*).

Once the sales agreement, which includes the product description and the purchaser information, has been approved by the sales department, it needs to be communicated to a shipment department that uses its own internal processes to facilitate the product shipment to the purchaser.

After the shipping department receives the sales agreement, it interacts with the transport-requirements registry to find the product transport requirements, and then it communicates concurrently with the insurance company to obtain insurance, and with the transporter to arrange the transport contract.

Insurance is obtained by invoking a smart contract method of the insurance company, while providing it with information on sales agreement that includes information on the product to be shipped, shipment destination, manner of transport, etc. Obtaining a transporter is achieved in a similar manner by invoking a smart contract method. The transporter responds by providing the contract for transport of the product.

Following this, the transporter performs the transport and when finished, the confirmation of delivery is provided by the transporter. Finally, once the product is delivered, payments are finalized.

If all interactions amongst the actors can be achieved by certified exchange of documents, the transformation of the BPMN model into the methods of a smart contract(s) can be achieved without requiring coding of task element scripts.

V. RELATED WORK

Closest to our research is the work on transforming BPMN models to smart contracts. The Lorikeet project [7] employs a two-phase methodology for converting BPMN models into smart contracts. First, the BPMN model is analyzed and transformed into smart contract methods, which are subsequently deployed and executed on a blockchain platform, specifically Ethereum. An off-chain component handles communication with the decentralized application (DApp), ensuring that actors exchange messages according to the BPMN model. The project also supports asset control, including both fungible and non-fungible tokens, and provides a registry and management methods for assets, such as transfers.

Caterpillar [5, 6] adopts a different approach by focusing on BPMN models confined within a single pool (a BPMN construct) where all business processes are recorded on the blockchain. Its architecture consists of three layers: Web Portal, Off-chain Runtime, and On-chain Runtime. The On-chain Runtime layer includes smart contracts for workflow control, interaction management, configuration, and process management, with Ethereum as the preferred blockchain platform.

Loukil et al. (2021) [9] proposed CoBuP, a collaborative business process execution architecture on blockchain. Unlike other methodologies, CoBuP does not directly compile BPMN models into smart contracts. Instead, it deploys a generic smart contract that invokes predefined functions. CoBuP's three-layer architecture, comprising Conceptual, Data, and Flow layers, transforms BPMN models into a JSON Workflow model that governs the execution of process instances, which in turn interact with data structures on the blockchain.

Similar to CoBuP, Bagozi et al. [34] employ a three-layer approach, albeit in a simpler form. In the first layer, a business analyst represents the collaborative process in BPMN. In the second layer, a business expert annotates the BPMN model to identify trust-demanding objects, after which Abstract Smart Contracts, independent of any specific blockchain technology, are created. Finally, Concrete Smart Contracts are generated and deployed on a specific blockchain platform.

## VI. SUMMARY AND CONCLUSIONS

In this paper we described how we are modifying our tool TABS+R to facilitate generation of smart contracts only under the guidance of a BA without assistance by a software developer – albeit, only for applications that interact only by exchange of documents. This is achieved by off-loading computation, performed by the script/code of task elements, off-chain and by facilitating collaboration of actors through a certified exchange of documents. We are thus facilitating democratization of the blockchain smart contracts by reducing the need for software development expertise.

Secondly, we show that generation of smart contracts using the TABS+R approach and its tool is flexible in that it can be used not only by large companies with sophisticated IT, but also by SMEs without IT to support software development. Although we developed the concept and supporting tool showing the feasibility of the approach, actual success can only be achieved by further experimentation and in particular development of complementary tools to achieve user experience that is expected in use of commercial software.

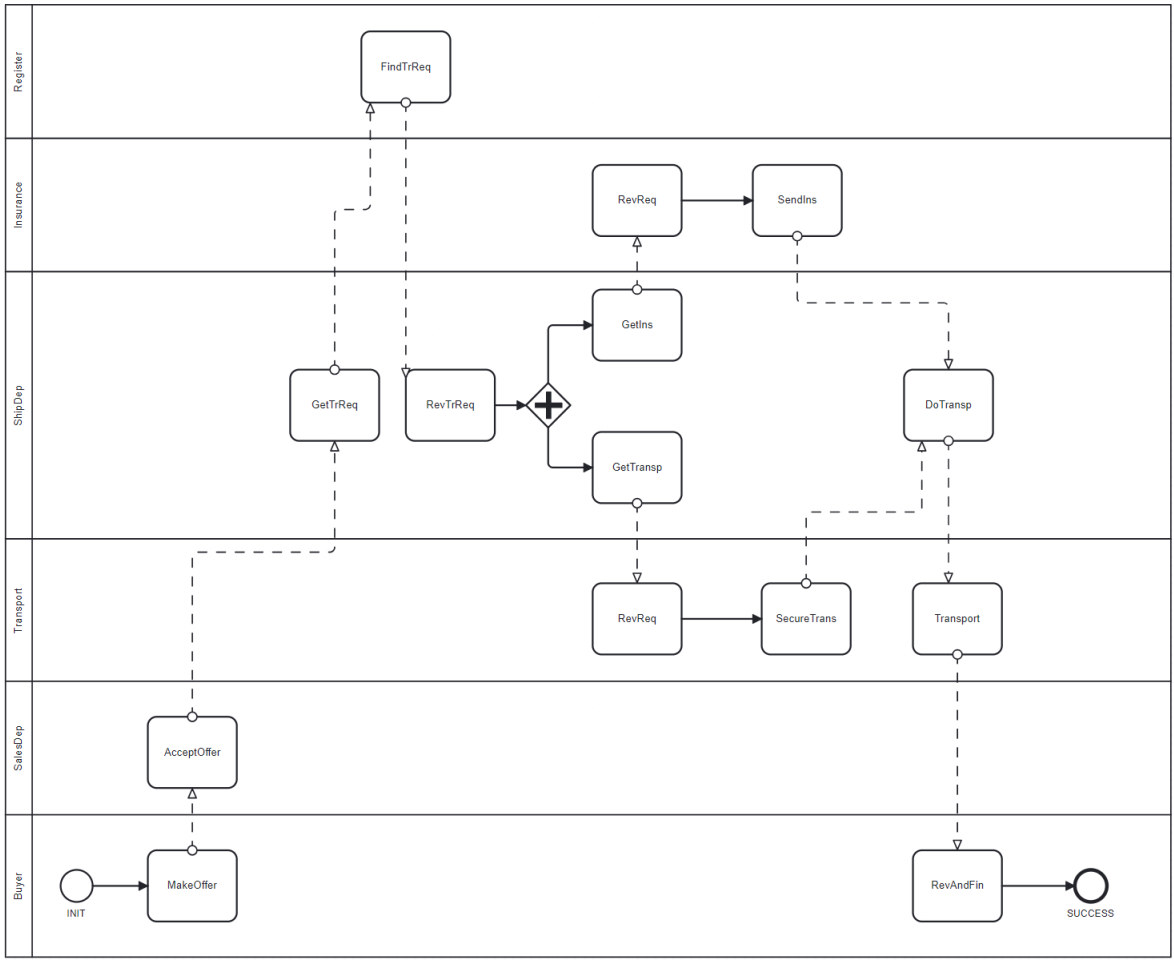

Figure 2. BPMN Model in the Context of a Large Company